%Paper: astro-ph/9403052
%From: narayan@surya.harvard.edu (Ramesh Narayan)
%Date: Wed, 23 Mar 94 10:13:33 EST

% The paper has been accepted for publication in:
%    The Astrophysical Journal (Letters)
% The text of the paper is in plain TeX.
% Figure 1 is attached as a postscript file at the end.
%    Cut and print separately.

\magnification\magstep1
\def\ssp{\baselineskip=11pt plus 1pt minus 1pt}

\def\ref{\medskip\hangindent=1cm\hangafter=1 \noindent}
\def\sles{\lower2pt\hbox{$\buildrel {\scriptstyle <}
   \over {\scriptstyle\sim}$}}
\def\sgreat{\lower2pt\hbox{$\buildrel {\scriptstyle >}
   \over {\scriptstyle\sim}$}}

\ssp
\centerline{\bf ADVECTION-DOMINATED ACCRETION: A SELF-SIMILAR SOLUTION}
\bigskip\bigskip
\centerline{\it Ramesh Narayan and Insu Yi}
\bigskip
\centerline{Harvard-Smithsonian Center for Astrophysics,}
\centerline{60 Garden Street, Cambridge, MA 02138}
\bigskip\bigskip
\noindent

\noindent{\bf Abstract} \bigskip We consider viscous rotating
accretion flows in which most of the viscously dissipated energy is
stored as entropy rather than being radiated.  Such {\it
advection-dominated} flows may occur when the optical depth is either
very small or very large.  We obtain a family of self-similar
solutions where the temperature of the accreting gas is nearly virial
and the flow is quasi-spherical.  The gas rotates at much
less than the Keplerian angular velocity; therefore, the central stars
in such flows will cease to spin up long before they reach the
break-up limit.  Further, the Bernoulli parameter is positive,
implying that advection-dominated flows are susceptible to producing
outflows.  Convection is likely in many of these flows and, if
present, will tend to enhance the above effects.  We suggest that
advection-dominated accretion may provide an explanation for the slow
spin rates of accreting stars and the widespread occurrence of
outflows and jets in accreting systems.

\bigskip\noindent
{\it Subject headings}:
accretion: accretion disks---black hole physics---hydrodynamics

%\vfill\eject
\bigskip\bigskip
\noindent{\bf 1. Introduction}
\bigskip

In astrophysical accretion flows, gravitational potential energy is
converted to kinetic and thermal energy of the accreting gas.  If the
thermal energy is efficiently radiated away, the orbiting gas becomes
much cooler than the local virial temperature and takes up a thin
disk-like configuration.  Accretion disk solutions with these
characteristics, derived originally by Shakura \& Sunyaev (1973) and
Lynden-Bell \& Pringle (1974), have been used to model a variety of
accreting systems in astrophysics (see Frank, King, \& Raine 1992 for
a review).

We ask here a simple question: What is the nature of the flow if the
accreting gas is unable to cool efficiently?  We use the term {\it
advection-dominated} to refer to a flow where the bulk of the
liberated thermal energy is carried in by the accreting gas as entropy
rather than being radiated.  In this {\it Letter} we write
height-averaged equations to describe advection-dominated flows and
discuss the properties of a particular class of self-similar
solutions.

\bigskip\bigskip
\noindent{\bf 2. Height-Averaged Equations of Advection-Dominated Accretion}
\bigskip

As in the standard theory of thin accretion disks, we vertically
average the flow equations and consider a two-dimensional flow in the
equatorial $R\phi$ plane.  We assume a steady axisymmetric flow so
that $\partial/\partial t=\partial/\partial\phi=0$ and all flow
variables are functions only of $R$.

Assuming gas pressure to dominate, we write the pressure as $P=\rho
c_s^2$ where $\rho (R)$ is the height-averaged density and $c_s^2(R)$
is the isothermal sound speed.  We denote the Keplerian angular
velocity by $\Omega_K(R)=(GM/R^3)^{1/2}$ and define the Keplerian
velocity $v_K=(GM/R)^{1/2}$, where $M$ is the central mass.  The
surface density of the gas is $\Sigma=2\rho H$, where $H\sim Rc_s/v_K$
is the vertical scale height.  Following Shakura \& Sunyaev (1973) we
take the kinematic coefficient of shear viscosity to be $\nu=\alpha
c_s H=\alpha c_s^2/\Omega_K$, where $\alpha$ is a constant.  Purely on
dimensional grounds, this approximation seems to be particularly
appropriate for the self-similar flows discussed here.

The density of the gas $\rho$, its radial
velocity $v$, angular velocity $\Omega$, and isothermal sound speed
$c_s$, satisfy the following four differential equations, namely the
continuity equation, the radial and azimuthal components of the
momentum equation, and the energy equation (e.g. Abramowicz et al.
1988, Narayan \& Popham 1993):
$$
{d\over dR}(\rho RH v)=0, \eqno (1)
$$
$$
v{dv\over
dR}-\Omega^2R=-\Omega_{K}^2R-{1\over\rho} {d\over dR}(\rho c_s^2),
\eqno (2)
$$
$$
v{d(\Omega R^2)\over dR}={1\over \rho RH}
{d\over dR}\left({\alpha \rho c_s^2 R^3H \over \Omega_{K}}
{d\Omega\over dR}\right), \eqno (3)
$$
$$
\Sigma vT{ds\over dR}=
{3+3\epsilon\over 2}2\rho H
v{dc_s^2 \over dR}-2c_s^2 Hv{d\rho\over dR}=Q^+-Q^-.\eqno (4)
$$
In equation (4), the left-hand side is the advected entropy,
where $T$ is the temperature and $s$ the entropy,
while the right-hand side gives the difference between the
energy input per unit area due to viscous dissipation ($Q^+$) and
the energy loss through radiative cooling ($Q^-$).  For convenience we have
defined a parameter $\epsilon= (5/3-\gamma)/(\gamma-1)$, where
$\gamma$ is the ratio of specific heats; note that $\epsilon=0$ in the limit
$\gamma=5/3$ and $\epsilon=1$ when $\gamma=4/3$.  Substituting
the viscous dissipation rate for $Q^+$, we obtain
$$
Q^+-Q^-= {2\alpha \rho c_s^2 R^2H
\over \Omega_{K}}\left(d\Omega\over dR\right)^2- Q^{-}
\equiv f{2\alpha \rho c_s^2 R^2H
\over \Omega_{K}}\left(d\Omega\over dR\right)^2. \eqno (5)
$$
The parameter $f$ measures the degree to which the flow is
advection-dominated.  In the extreme limit of no radiative cooling, we
have $f=1$, while in the opposite limit of very efficient cooling,
$f=0$.  Finally, we define $\epsilon^{'}\equiv\epsilon/f$.  This is an
important parameter which plays a critical role in determining the
nature of the flow.

Let us for simplicity assume that $\epsilon^{'}$
is independent of $R$.
Equations (1)--(4) then permit a self-similar solution of the form
$$
\rho\propto R^{-3/2},\qquad
v\propto R^{-1/2},\qquad
\Omega\propto R^{-3/2},\qquad
c_s^2\propto R^{-1}, \eqno (6)
$$
where
$$
v=-(5+2\epsilon^{'}){g(\alpha,\epsilon^{'})\over3\alpha}
v_K\approx -{3\alpha\over (5+2\epsilon^{'})}v_K, \eqno (7)
$$
$$
\Omega=\left[{2
\epsilon^{'}(5+2\epsilon^{'})g(\alpha,\epsilon^{'})
\over 9\alpha^2}\right]^{1/2}
\Omega_K\approx \left[{2\epsilon^{'}\over
5+2\epsilon^{'}}\right]^{1/2}\Omega_K, \eqno (8)
$$
$$
c_s^2={2(5+2\epsilon^{'})\over 9}{g(\alpha,\epsilon^{'})\over\alpha^2}v_K^2
\approx {2\over 5+2\epsilon^{'}}v_K^2, \eqno (9)
$$
$$
g(\alpha,\epsilon^{'})\equiv \left[{1+{18\alpha^2\over (5+2\epsilon^{'})^2}}
\right]^{1/2}-1. \eqno (10)
$$
The density $\rho$ may be obtained from the mass accretion rate, $\dot
M=-4\pi RHv\rho$.  In equations (7)--(9), the first relation gives the
exact solution, while the second corresponds to the limit when
$\alpha^2\ll 1$.  If we include bulk viscosity through an $\alpha$
prescription there are corrections of order $\alpha^2$ in the results.
Also, if $\epsilon<0$, i.e. if $\gamma>5/3$, there is a second class
of self-similar rotating {\it wind} solutions where $v>0$.  We do not
discuss these extensions here.

\bigskip\bigskip
\noindent{\bf 3. Properties of the Self-Similar Solution}
\bigskip

In the limit of very efficient cooling, $f\rightarrow0$ and
$\epsilon^{'}\rightarrow\infty$, and the solution given in equations
(7)--(9) corresponds to a standard thin accretion disk with $v,c_s\ll
v_K$ and $\Omega\rightarrow\Omega_K$.  The properties of such disks
have been widely discussed in the literature.  In this {\it Letter} we
are interested in the opposite limit of advection-dominated flows
where $f$ is a reasonable fraction of unity and
$\epsilon^{'}\sim\epsilon<1$.  In this limit, the self-similar
solution derived above has several interesting properties which we now
discuss.  Although our discussion is based on this particular
solution, we believe that the results are valid for general
advection-dominated accretion flows.

Equation (9) shows that the sound speed in advection-dominated flows
is comparable to the Keplerian speed $v_K$, which means that the
temperature of the accreting gas is nearly virial.  This is of course
expected since by assumption the gas has no way to cool.  When
$c_s\sim v_K$, the disk vertical thickness $H$ is comparable to $R$
and the flow is quasi-spherical.  Unfortunately, the vertical
averaging on which equations (1)--(4) are based is suspect when
$H/R\sim 1$, and therefore some of the numerical factors in our
results may be in error.

The radial velocity of the accreting gas is proportional to $\alpha$.
This is because the radial speed is determined principally by how fast
the viscosity can move angular momentum outwards.  Since $v\sim\alpha
c_s^2/v_K$, the radial velocity tends to be much larger in
advection-dominated flows than in thin disks.

When $\epsilon^{'}$ is small, we see that the angular velocity
$\Omega$ of the flow is smaller than the local Keplerian $\Omega_K$ by
a factor $\sim(\epsilon^{'})^{1/2}$.  Indeed, as
$\epsilon\rightarrow0$, which corresponds to $\gamma\rightarrow 5/3$,
$\Omega$ goes to zero and our solution matches on to the Bondi
spherical accretion solution for this $\gamma$.  (In a sense, the
solution (7)--(9) is a natural extension of self-similar $\gamma=5/3$
Bondi accretion to rotating flows and general $\gamma$.)  Because
advection-dominated accretion has a significantly sub-Keplerian
$\Omega$, if such a flow were to be present around an accreting star,
the star would cease spinning up long before the rotation rate
approached the break-up limit.  This may have important implications
for the spin rate of accreting stars.

Equation (2) shows that the self-similar solution
satisfies the relation
$$
{1\over 2}v^2+\Omega^2R^2-\Omega_K^2R^2+{5\over 2}c_s^2=0. \eqno (11)
$$
{}From this we can compute the normalized parameter $b\equiv Be/v_K^2$,
where $Be$ is the Bernoulli constant:
$$
b={1\over v_K^2}\left({1\over 2}v^2+{1\over 2}\Omega^2R^2-\Omega_K^2R^2+
{\gamma\over\gamma-1}c_s^2\right)
$$
$$
=-{\Omega^2R^2\over 2v_K^2}+\left ({\gamma\over\gamma-1}-{5\over2}
\right ){c_s^2\over v_K^2}={3\epsilon-\epsilon^{'}
\over 5+2\epsilon^{'}}. \eqno (12)
$$
As is well-known, $Be$ is conserved in adiabatic flows in the absence
of viscosity.  Therefore, whenever $b$ is positive, it implies that if
we were somehow to turn the gas around and to let it flow
adiabatically outward on a radial trajectory, the gas would reach
infinity with a net positive energy.  On the other hand, if $b<0$, the
gas cannot spontaneously escape to infinity.  Equation (12) shows that
$b$ is positive in advection-dominated flows for all values of
$\alpha$ and for any $\gamma<5/3$ so long as $f>1/3$.  Winds, jets and
other outflows are common in several accreting systems in astrophysics
and it is tempting to speculate that these may originate in
advection-dominated flows.

Note that the positivity of the Bernoulli constant does not imply a
lack of conservation of energy, but simply arises because viscous
stresses transfer energy from small to large radii.  Even in the
standard theory of thin accretion disks, the energy radiated from any
given annulus of the disk is greater by a factor of 3 than the net
gravitational energy released within that annulus (e.g.  Frank et al.
1992).  A similar phenomenon occurs in our solutions, and the excess
$b$ at any radius represents energy transferred to that radius from
smaller radii.  When proper boundary conditions are applied and the
full global problem is solved, total energy will of course be
conserved.

It is easily shown that the entropy increases inwards in our
self-similar solution.  We need therefore to consider the possibility
of convective energy transport.  In a rotating medium, the condition
for a dynamical convective instability is
$$
N_{eff}^2\equiv N^2+\kappa^2=-{1\over\rho}{dP\over dR}
{d\ln s\over dR}+\kappa^2<0,\eqno (13)
$$
where $N$ is the usual Brunt-V\"ais\"al\"a frequency and $\kappa$ is
the epicyclic frequency which in our case is equal to $\Omega$.
For our self-similar solution we find (in the limit when $\alpha^2\ll1$)
$$
N_{eff}^2={10\epsilon^{'}+6\epsilon\epsilon^{'}-15\epsilon
\over(5+3\epsilon)(5+2\epsilon^{'})}\Omega_K^2.\eqno (14)
$$
This gives a dynamical instability for
$$
f>{2\over3}+{2\over5}\epsilon.\eqno (15)
$$
Even when $f$ is below this limit a mild double diffusive
instability will be present.

Although advection-dominated flows are convectively unstable, we can show that
convection, if at all, only strengthens our main conclusions.  The
height-integrated convective energy flux $F_c$ in the presence of
an unstable entropy gradient can be written quite generally as
$$
F_c=-\Sigma K_cT{ds\over dR},\eqno (16)
$$
where $K_c$ is the diffusion constant associated with convective
transport.  In the spirit of the Shakura-Sunyaev viscosity formula let
us write $K_c=\alpha_cc_s^2/\Omega_K$.  Substituting our self-similar
solution into equation (16) we then find that $F_c$ scales as $R^{-2}$
and has a negative divergence.  Convection therefore provides an
additional energy source in equation (4) of magnitude
$$
Q_c^+=-\vec\nabla\cdot\vec F=-\Sigma{\alpha_cc_s^2\over v_K}
T{ds\over dR}.\eqno (17)
$$
This term has the same structure as the
left-hand side of equation (4).  We
can therefore combine the two terms by simply redefining $\epsilon^{'}$
to be
$$
\epsilon^{'}={\epsilon\over f}\left(1+{\alpha_cc_s^2\over vv_K}
\right)={\epsilon\over f}\left(1-{2\over3}{\alpha_c\over\alpha}
\right).\eqno (18)
$$
With this definition, all of the results in equations (6)--(10)
continue to hold.

In general we expect $\alpha\ \sgreat\ \alpha_c$.  This is because
convective turbulence contributes to the viscosity but not all sources
of viscosity produce bulk energy transport.  (Ryu \& Goodman (1992)
show that if the entropy in an accretion disk is stratified
vertically, then convection actually moves angular momentum inwards;
in our case we have a radially stratified medium and so we expect
convection to produce a normal outwardly-oriented angular momentum
flux.)  Since $\alpha\ \sgreat\ \alpha_c$, equation (18) shows that
convection only introduces a modest perturbation to the energy
equation (4).  This is very different from the situation in stellar
interiors where convection, once it sets in, strongly modifies the
structure and essentially forces an isentropic radial distribution.
In our case, the strong advection ensures that convection can only be
a moderate perturbation.  Indeed, as equation (18) shows,
the primary effect of convection is to {\it reduce} the value of the
parameter $\epsilon^{'}$.  This means that the various properties of
advection-dominated flows discussed earlier become even more
pronounced.  In particular, when there is convection, the
parameter $b$ remains positive for a wider range of $f$ than in the
non-convective case.  The reason is that convection transports energy
from smaller to larger $R$ and therefore adds to the effect of the
viscous stress.

The above discussion suggests that the main results of this paper are
likely to survive even if there are other instabilities in the flow.

\bigskip\bigskip
\noindent{\bf 4. Discussion}
\bigskip

Advection-dominated accretion is expected to occur in several
astrophysical situations.  Thin accretion disks for example exhibit a
thermal instability at sufficiently low optical depth, when the
cooling through free-free emission is unable to keep up with the
viscous energy generation.  Numerical models of accretion disks in
cataclysmic variables (Narayan \& Popham 1993) show that, under the
influence of this instability, the disk switches to an
advection-dominated flow where only a fraction of the released energy
is radiated.  A more extreme possibility is that at very low accretion
rates the infalling material may never cool sufficiently to collapse
to a thin disk, and we could imagine an advection-dominated flow all
the way from the outermost radius down to the central star or black
hole.

Advection can dominate also in the opposite limit when the mass
accretion rate is very high and the optical depth $\tau$ of the disk
is large.  If the cooling time of the disk, $t_{cool}\sim H\tau/c$, is
longer than the accretion time $R/v$, most of the accretion energy is
retained in the gas and we have an advection-dominated flow.  Such
conditions are possible in young stellar objects and symbiotic stars
in outburst (Popham et al. 1993) and in accreting neutron stars and
black holes at high mass accretion rates.

Thus, advection-dominated accretion is likely to occur in a number of
astrophysical systems, and some objects may have both
advection-dominated and standard cooling-dominated zones at different
radii.  Surprisingly, very few studies in the literature have included
advection effects, and the physics of advection-dominated accretion
has been hardly discussed at all.  The self-similar solutions we
present in \S 2 of this {\it Letter} represent a first step towards
understanding such flows.

Our self-similar solutions have some interesting features which we have
described in \S 3.  Here we would like to highlight the following:

\item{1.} The angular velocity $\Omega$ of the flow is
less than the Keplerian angular velocity $\Omega_K$; in fact, for the
astrophysically interesting case of $\gamma\rightarrow 5/3$
($\epsilon\rightarrow0$), we find $\Omega\ll\Omega_K$.  In accretion
theory it is usually assumed that the infalling material rotates with
$\Omega\sim\Omega_K$ and that accreting stars will spin up to the
``break-up limit'' corresponding to a stellar spin rate
$\Omega_*\sim\Omega_K(R_*)$, where $R_*$ is the equatorial radius of
the star.  Advection-dominated accretion flows are very different and
the star may reach an equilibrium spin state with
$\Omega_*\ll\Omega_K(R_*)$.  At the equilibrium $\Omega_*$ the system
will adjust the angular momentum flux in such a way that $\Omega_*$
remains constant (Popham \& Narayan 1991, Paczy\'nski 1991).
Similarly, in the case of accretion onto a magnetic star it is usually
assumed that the central star spins up until $\Omega_*=\Omega_K(R_A)$,
where $R_A$ is the Alfven radius.  Once again, if the accretion is
advection-dominated, we can have $\Omega_*\ll\Omega_K(R_A)$.

\item{2.} The scaled Bernoulli parameter $b$ (see equation 12)
is positive in self-similar advection-dominated flows for a wide range
of parameters.  We have argued that this means the gas in such flows
is capable of spontaneously escaping to infinity.  We suggest that
advection-dominated flows may provide a generic explanation for many
of the outflows and jets that are so ubiquitous in accreting systems.
It is beyond the scope of this {\it Letter} to discuss the exact
mechanism by which outflows may be generated.  One possibility is that
a shock may divert part of the high entropy accreting material into an
outward-pointing trajectory.  Alternatively, the positivity of $b$ may
make the generation of winds so easy that even a quite modest
radiative or magnetic stress may set off a substantial wind.  (We must
caution however that, in some circumstances, once the material reaches
the surface of the disk it may cool and lose its positive $b$, perhaps
making it harder to generate a wind.)

\item{3.} By definition, the radiative luminosity
of an accretion-dominated flow is much less than the standard $GM\dot
M/R_*$ that is usually associated with accretion.  Therefore,
estimates of the mass accretion rates of observed systems could be
seriously in error, especially when the central star is a black hole
which can swallow the advected entropy, or when much of the accretion
energy escapes as an outflow.  Moreover, since the accreting gas is
almost at virial temperature, the spectrum is likely to be much hotter
than the spectrum of an equivalent thin disk.  This is particularly
true of optically thin advection-dominated flows.

\item{4.} Advection-dominated flows are convectively unstable.
Because convection transfers energy from small to large radii, we find
that it enhances the effects described above.

The self-similar solutions described here technically extend from
$R=0$ to $R=\infty$ with no boundaries.  How relevant are the results
if the accretion flow is advection-dominated only over a finite range
of $R$ and is bounded on the outside and inside by non-advection
zones?  We have carried out some preliminary investigations of this
question by numerically solving equations (1)--(4) for various
boundary conditions.  Figure 1 shows an example where the accreting
gas starts off as a thin disk at large $R$, goes through an
advection-dominated phase, and becomes a rotating settling ``star'' at
small $R$.  Over a range of intermediate radii the numerical solution
is close to the self-similar form.  We have calculated other numerical
solutions where the central star is a black hole and the flow goes
through a sonic point, and again we find that the flow is nearly
self-similar in between the two boundaries.

{}From these numerical examples we conclude that the self-similar
solutions we have described may be of more than academic interest and
that perhaps real flows often resemble the self-similar form.  In any
case, the main features of these solutions, namely slow rotation and
the ability to make outflows, are probably generic to a wide class of
advection-dominated flows.

\bigskip\noindent
{\it Acknowledgements}: RN thanks Bob Popham for numerous discussions
on the physics of accretion, advection and spin-up.  We are grateful
to Lee Hartmann for helpful comments and the referee for useful
criticism.  This work was supported in part by NSF grant AST 9148279.

%\vfill\eject
\bigskip\bigskip
\noindent{\bf References}
\bigskip
\ref{Abramowicz, M., Czerny, B., Lasota, J. P., \& Szuszkiewicz, E. 1988, ApJ,
332, 646}

\ref{Frank, J., King, A., \& Raine, D. 1992, Accretion Power in Astrophysics
(Cambridge, UK: Cambridge University Press)}

\ref{Lynden-Bell, D., \& Pringle, J. E. 1974, MNRAS, 168, 603}

\ref{Narayan, R., \& Popham, R. 1993, Nature, 362, 820}

\ref{Paczy\'nski, B. 1991, ApJ, 370, 597}

\ref{Popham, R., \& Narayan, R. 1991, ApJ, 370, 604}

\ref{Popham, R., Narayan, R., Hartmann, L., \& Kenyon, S. 1993, ApJ,
415, L127}

\ref{Ryu, D., \& Goodman, J. 1992, ApJ, 388, 438}

\ref{Shakura, N. I., \& Sunyaev, R. A. 1973, A\&A, 24, 337}

%\vfill\eject
\bigskip\bigskip
\noindent{\bf Figure Caption}
\bigskip\noindent
Figure 1. A numerical solution of equations (1)--(4) with $GM=1$,
$\alpha=0.1$, $f=1$, and $\epsilon=1/3$ (or $\gamma=1.5$).  At
$R=10^6$ (the outer edge), the boundary conditions are
$\Omega=\Omega_K$ and $c_s=0.001v_K$, appropriate to a very thin disk.
At $R=1$ (the inner edge), the boundary condition is $v=0.05c_s$,
corresponding to a settling star.  The ``star'' in this example
extends out to about $\log (R)\sim 1$.  The angular momentum flux has
been set to zero, corresponding approximately to an equilibrium spinup
state.  (The precise equilibrium value of this parameter depends on
the structure of the central star, cf Popham \& Narayan 1991).  The
solid lines in the three panels show the radial velocity $v$, the
scaled angular velocity $\Omega/\Omega_K$, and the scaled Bernoulli
parameter $b$.  The dashed lines show the results corresponding to the
self-similar solution.

\bye